\documentclass[aps, prb, twocolumn, footinbib, showkeys, superscriptaddress, floatfix]{revtex4-1}

\usepackage[utf8]{inputenc}
\usepackage[T1]{fontenc}

\usepackage[english]{babel}
\usepackage{amsmath, amssymb, bm, soul}

\usepackage[pdftex]{graphicx}
\usepackage[hypcap=true]{caption}
\usepackage[hypcap=true, labelformat=simple]{subcaption}
\usepackage{ragged2e}
\DeclareCaptionJustification{justified}{\justifying}
\usepackage{color}
\usepackage[usenames, dvipsnames]{xcolor}
\usepackage[linktoc=page, colorlinks=true, linkcolor=MidnightBlue, citecolor=MidnightBlue, urlcolor=OliveGreen]{hyperref} 

\setul{1.4pt}{.4pt}
\newcommand{\uhref}[2]{{\href{#1}{\ul{#2}}}}

\newcommand{\subfignum}[1]{{\color{MidnightBlue}#1}}

\usepackage{enumerate}
\usepackage{balance}

\newcommand{\Jc}{J_\mathrm{c}}
\newcommand{\Ic}{I_\mathrm{c}}
\newcommand{\Jcmax}{J_\mathrm{c,max}}
\newcommand{\Jdp}{J_\mathrm{dp}}
\newcommand{\Jl}{J_\mathrm{l}}
\newcommand{\Jcu}{J_\mathrm{c,u}}
\newcommand{\Jcright}{J_\mathrm{c\rightarrow}}
\newcommand{\Jcup}{J_\mathrm{c\uparrow}}
\newcommand{\Tc}{T_\mathrm{c}}
\newcommand{\Tci}{T_\mathrm{c,i}}
\newcommand{\ei}{\varepsilon_\mathrm{i}}
\newcommand{\Tcb}{T_\mathrm{c,b}}
\newcommand{\Tf}{T_\mathrm{f}}
\newcommand{\Ec}{E_\mathrm{c}}
\newcommand{\Hct}{H_\mathrm{c2}}
\newcommand{\Bm}{B_\Phi}
\newcommand{\Bmmax}{B_\mathrm{\Phi,max}}
\renewcommand{\i}{i}
\renewcommand{\r}{\boldsymbol{r}}
\newcommand{\A}{\boldsymbol{A}}
\newcommand{\B}{\boldsymbol{B}}
\newcommand{\J}{\boldsymbol{J}}
\renewcommand{\st}{\textsuperscript{st}}
\newcommand{\nd}{\textsuperscript{nd}}
\newcommand{\rd}{\textsuperscript{rd}}
\renewcommand{\th}{\textsuperscript{th}}

\renewcommand{\thesection}{\arabic{section}}

\begin{document}

\title{Targeted evolution of pinning landscapes for large superconducting critical currents}

\author{Ivan A. Sadovskyy}
\affiliation{Materials Science Division, Argonne National Laboratory, Lemont, IL 60439}

\author{Alexei E. Koshelev} 
\affiliation{Materials Science Division, Argonne National Laboratory, Lemont, IL 60439}

\author{Wai-Kwong Kwok}
\affiliation{Materials Science Division, Argonne National Laboratory, Lemont, IL 60439}

\author{Ulrich Welp}
\affiliation{Materials Science Division, Argonne National Laboratory, Lemont, IL 60439}

\author{Andreas Glatz}
\affiliation{Materials Science Division, Argonne National Laboratory, Lemont, IL 60439}
\affiliation{Department of Physics, Northern Illinois University, DeKalb, IL 60115}

\begin{abstract}
The ability of type-II superconductors to carry large amounts of current at high magnetic fields is a key requirement for future design innovations in high-field magnets for accelerators and compact fusion reactors and largely depends on the vortex pinning landscape comprised of material defects. The complex interaction of vortices with defects that can be grown chemically, e.g., self-assembled nanoparticles and nanorods, or introduced by post-synthesis particle irradiation precludes \textit{a priori} prediction of the critical current and can result in highly non-trivial effects on the critical current. Here, we borrow concepts from biological evolution to create a genetic algorithm evolving pinning landscapes to accommodate vortex pinning and determine the best possible configuration of inclusions for two different scenarios: an evolution process starting from a pristine system and one with pre-existing defects to demonstrate the potential for a post-processing approach to enhance critical currents. Furthermore, the presented approach is even more general and can be adapted to address various other targeted material optimization problems.
\end{abstract}


\maketitle

\tableofcontents

\section{Introduction}

Life has undergone tremendous changes due to natural selection~--- from relatively simple molecules with replication capability to complex organisms, whose understanding is still far beyond present contemplation. Modern computer systems have enabled the effective exploitation of the idea of natural selection for practical purposes. The underlying genetic algorithms are widely used in electromagnetic\cite{Weile:1997} and mechanical design (e.g., the design of space antenna\cite{Hornby:2015}), financial mathematics, energy applications, scheduling problems, circuit design, image processing, medicine, etc. Within this approach, one only needs to specify the direction of positive mutations in order to find optimal or beneficial characteristics of the system of interest, i.e., replace natural evolution by \textit{targeted evolution}, which is especially effective in complex systems with a large number of degrees of freedom.

A key science aspect to advance the deployment of high-temperature superconductors (HTSs) is the discovery of novel materials which can carry large currents without dissipation at high magnetic fields.\cite{Campbell:1972} These materials are especially desirable for high-performance applications\cite{Kwok:2016} such as superconducting motors, generators, magnets, and power lines in urban areas.\cite{Malozemoff:2012, Senatore:2014, Obradors:2014, Shiohara:2012, Lee:2008} Low dissipation is also very important for superconducting cavities for particle accelerators,\cite{Padamsee:2014, Gurevich:2017} undulators for X-ray synchrotrons,\cite{Kesgin:2017} and compact fusion reactors.\cite{Kramer:2015} The main challenge is to suppress the dissipation in these systems caused by the motion of quantized elastic magnetic flux tubes or vortices, which appear in type-II superconductors in magnetic fields above the first critical field.\cite{Abrikosov:1957} Since most applied superconductors are of type-II, the study of efficient pinning mechanisms benefits a majority of superconducting technologies. Vortices can be trapped (or pinned) by inhomogeneities in the material, usually in the form of non-superconducting defects.\cite{Blatter:1994, Brandt:1995, Larkin:1979} Examples are point-like pinning centers (impurities, vacancies, inclusions), one-dimensional defects (dislocations, irradiation tracks), or two-dimensional defects (twin boundaries, stacking faults). Although extensive knowledge has been gained in the pursuit of high critical currents (the highest current the system can carry without dissipation),\cite{Kramer:1973, DewHughes:1974, Kes:1992, Scanlan:2004, Godeke:2006, Foltyn:2007} the fundamental solution to the dynamics of interacting vortices in disordered media is still unknown. Only recently more systematic, computer-assisted approaches were developed,\cite{Koshelev:2016, Sadovskyy:2016real} leading to the critical-current-by-design methodology.\cite{Sadovskyy:2016jc}

\begin{figure*}
	\centering \includegraphics[width=13.2cm]{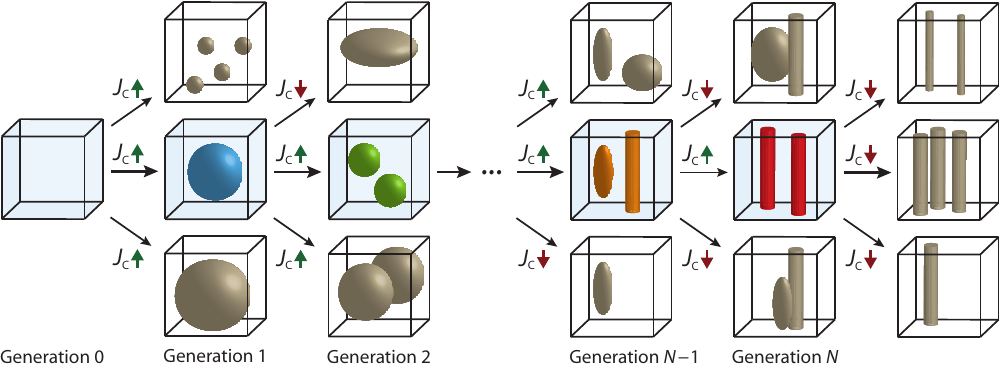}
	\caption{ \label{fig:evolution_sketch}
		Sketch of a targeted evolution of the pinning landscape. We start with generation~0, 
		which contains a single configuration without defects. Each defect has elliptical shape 
		and is characterized by three independent diameters. The evolution process, 
		`mutates' the pinning landscape by adding/removing, translating, scaling, and 
		reshaping particles. These mutations create the next generation. We accept the pinning 
		landscape with maximal critical current density, $\Jc$, and discard all others. 
		The evolution ends at some generation $N$ with configuration having 
		maximal $\Jc$ (shown in red).
	}
\end{figure*}

While sophisticated numerical optimization methods\cite{Kimmel:2017a} and corresponding experiments can guide the design of superconductors with enhanced critical current densities, $\Jc$, the problem requires defining the general geometry of the vortex pinning landscape (or pinscape) \textit{a priori}. This works well if only a certain type of pinning defects is present~--- in other words, pinscapes defined by only a few parameters. Hence, the overall best pinscape for the highest $\Jc$ cannot be determined by these approaches. To address this question, one needs to study all possible combinations of defects, resulting in highly mixed pinscapes. Each of the individual defects are described by numerous material and geometrical parameters, resulting in an extremely high-dimensional parameter space for the pinscape. This is where a genetic approach can be utilized.

In this work, we borrow concepts from biological evolution to create a vortex pinning genome with targeted evolution for predicting high in-field critical currents. We focus on the geometrical aspect of the defects to produce the best pinscape for a given system. In particular, we evolve the pinscape by changing the shapes of individual defects (see sketch of targeted evolution in Fig.~\ref{fig:evolution_sketch}), thereby including the possibility of all major defect types such as columnar and spherical defects, which can be experimentally realized. Moreover, our approach can also be adapted for many different materials optimization/design problems. Here, we demonstrate its power for (i)~numerically determining the maximum possible critical current density in superconductors with non-magnetic normal inclusions and (ii)~developing a universal post-processing strategy for enhancing the performance of superconductors with preexisting pinning landscapes such as in commercial HTS wires and superconductors in alternating or non-homogeneous magnetic fields.

\section{Targeted evolution}

An essential ingredient for our approach is to obtain the critical current for a given `evolved' pinscape. Here we describe the complex dynamics and pinning of vortices by the time-dependent Ginzburg-Landau (TDGL) equation,\cite{Schmid:1966} allowing us to determine the critical current density,\cite{Koshelev:2016,Sadovskyy:2016real} $\Jc$, --- the fitness function of the system~--- and to obtain detailed information about the vortex matter in bulk superconductors.\cite{Sadovskyy:2017hex, Willa:2018} The TDGL equation yields $\Jc$ as function of shapes, sizes, and positions of pinning defects, see details in Sec.~\ref{sec:tdgl}.

\begin{figure*}
	\begin{subfigure}[t]{0\linewidth} \phantomcaption \label{fig:evolution_snapshots}\end{subfigure}%
	\begin{subfigure}[t]{0\linewidth} \phantomcaption \label{fig:evolution_final_snapshot} \end{subfigure}%
	\begin{subfigure}[t]{\linewidth} \includegraphics[width=18cm]{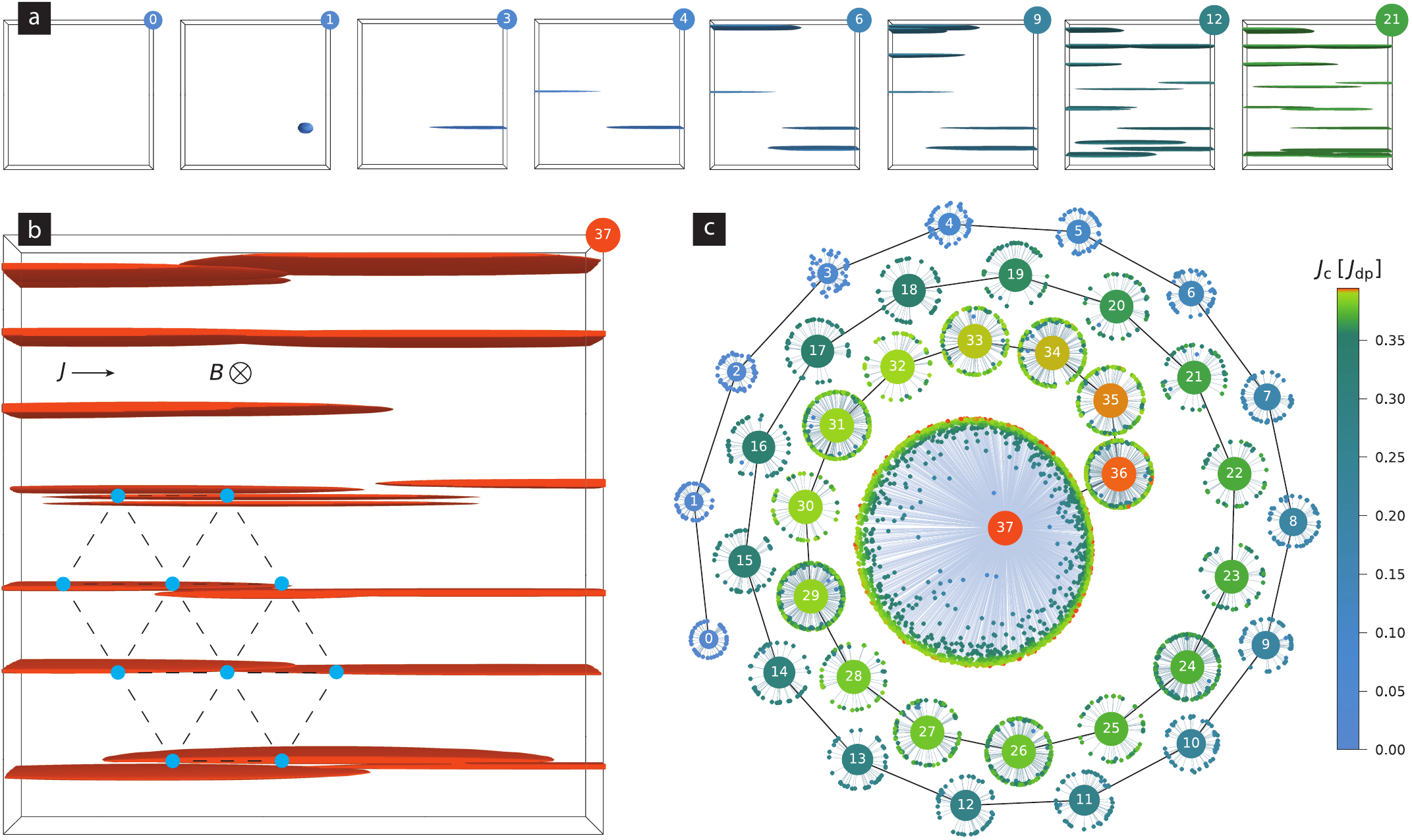} \phantomcaption \label{fig:evolution_path} \end{subfigure}
	\caption{ \label{fig:evolution}
		Evolution history. 
		\subref{fig:evolution_snapshots}~The evolution process starts with a superconductor without inclusions shown 
		in the left panel. The following panels show pinning landscapes having highest critical current, 
		$\Jc$, in {1\st}, {3\rd}, {4\th}, {6\th}, {9\th}, {12\th}, and {21\st}~generation, correspondingly. 
		In the {1\st}~generation the maximal critical current is achieved with the configuration containing 
		a single nearly spherical inclusion. In {2\nd} and {3\rd}~generations this inclusion evolves 
		to a flattened ellipsoid lying in the plane spanned by the current and magnetic field. 
		In subsequent generations, this ellipsoid is copied multiple times to enhance the total pinning. 
		The remaining generations of the evolution process `fine tunes' the landscape by copying, 
		removing, moving, and slightly deforming successors of the seed inclusion. 
		All steps are presented in a \uhref{https://youtu.be/2sKaJV-cLMM}{movie clip}.
		\subref{fig:evolution_final_snapshot}~The final pinning landscape consisting of a periodic array of almost 
		planar defects has the best possible critical current in the framework of our model. 
		The positions of pinned vortices are shown schematically by blue circles. 
		\subref{fig:evolution_path}~The evolution tree. The numbered circles represent configurations 
		with the maximum critical current per generation. Small dots around each numbered circle 
		are its successors, screened by targeted selection. The color of each pinning configuration 
		corresponds to its critical current shown by the color bar on the right.
	}
\end{figure*}

As a quite general model, we consider pinning landscapes containing $D$ ellipsoidal metallic pinning centers with principle axes ($a_i, b_i, c_i$), aligned in the $x$, $y$, $z$ directions, with center positions ($x_i, y_i, z_i$), where $i = 1, \ldots, D$. These ellipsoidal defects can describe a large variety of defect geometries in superconductors such as precipitates, point defects, dislocations, grain boundaries, and stacking faults, as well as particle-irradiation-induced columnar or spherical defects. For example, point defects can be modeled by small spherical inclusions, grain boundaries by flattened spheroids and columnar defects by spheroids with one of the diameters larger than the system size. To find pinscapes with ellipsoidal defects that yield the highest critical current, we employ an evolution-based algorithm with three distinct stages: (1)~mutations and targeted selection, (2)~extrapolation and analysis, (3)~and verification, described below. 

\textit{Stage 1: Mutations and targeted selection.} This step implements the evolutionary paradigm, during which the shape and position of individual inclusions is altered independently (mutation) and the critical current is calculated. A set of random mutations produces a new generation. Each new pinscape or successor may contain one (typical) or more sequential mutations (rare). Each pinscape is evaluated and the one with largest critical current is chosen for further evolution, see sketch in Fig.~\ref{fig:evolution_sketch}. The initial pinscape usually depends on the problem to be studied. For a general situation (discussed in the next section) one can initiate the targeted evolution algorithm with an empty pinscape, the `{0\th}~generation', which represents a homogeneous system with zero critical current. 

Mutations have random type, strength, direction, and number of affected inclusions, namely: 
\begin{itemize}
\item[(i)] copying of existing inclusions or adding new inclusions of random shape, 
\item[(ii)] removing inclusions, 
\item[(iii)] changing the inclusion principle axes $a_i$, $b_i$, and $c_i$,
\item[(iv)] changing the inclusion position ($x_i, y_i, z_i$), 
\item[(v)] repelling/attracting pairs of inclusions, i.e., increasing/decreasing the distance between randomly chosen inclusions $i$ and $j$, 
\item[(vi)] squishing inclusions, i.e., changing the inclusion's axes $a_i$, $b_i$, and $c_i$ while maintaining its volume, 
\item[(vii)] splitting inclusions, i.e., creating a pair of inclusions with the same volume as the original one, and 
\item[(viii)] merging pairs of inclusions.
\end{itemize}
Mutation types (vi)--(viii) preserve the volume of the affected inclusions. Note, that if we start the mutation process with an empty pinscape, the only possible mutation is the addition of defects. The set of mutated pinscapes represents a new generation. We calculate the critical current for each pinscape in a generation. These are then compared to the maximal critical current of the previous generation. In case none of the mutations increase the critical current, we repeat the mutation procedure and expand the population in the current generation until at least one pinscape produces a critical current larger than the maximal critical current of the previous generation or a maximum population is reached. This stage is implemented to work in parallel. If a configuration with larger critical current is found within a generation, we select the pinscape with largest critical current as seed configuration for the following generation and then apply the mutation procedure again. Repeating this protocol produces subsequent generations of pinscapes with even higher critical currents. We stop in generation~$N$ if no further critical current enhancement is found (the cutoff population size is 2048).

The evolution approach provides us with the types and parameters of defects that ensure maximum vortex pinning and, consequently, maximum critical current. The results are obtained without any assumptions of the pinscape structure, and only depend on external parameters such as magnetic field and temperature. In some application-relevant situations, the initial pinscape and the type of possible mutations may have some constraints, see Sec.~\ref{sec:applications}.

\textit{Stage 2: Extrapolation and analysis.} Stage 1 provides information regarding the distribution of the particle sizes and, in some cases, their spatial distribution. We can model/extrapolate these distributions with only a few parameters such as the size and typical distances between defects. In other words, one can use the general knowledge of the defect shapes obtained by the evolutionary approach and characterize the corresponding pinscape with a simplified global parameter set. For example, if the optimal pinscape consists of randomly distributed spherical defects of similar diameters, the configuration can be characterized by two parameters: concentration and diameter of the defects.\cite{Koshelev:2016}

Based on the simplified global parameter set, near-optimal pinscapes can be fine-tuned using conventional optimization methods.\cite{Kimmel:2017a} Furthermore, one can sample critical currents for near-optimal parameter sets to determine the robustness of the configuration, and compare them to analytical results.\cite{Blatter:2004, Nelson:1993, Blatter:1994, Willa:2016, Thomann:2012}

\textit{Stage 3: Verification.} To test the model obtained in stage 2, we restart the evolution process with the best model configuration and change the positions and sizes of each inclusion individually. The model is verified, if subsequent evolution cannot further increase the critical current by a significant amount (we typically use a threshold of 3\% within 2048 mutations).

Stages 2 and 3 are in a sense optional, as they elucidate the underlying mechanism for the optimal pinscape, extract a model, and show the stability of the process. Stage 1 alone can determine the general optimal pinscapes.

\section{Optimal pinscape in fixed field}

Starting with empty pinscapes and allowing almost any possible mutation is typically difficult to realize in commercial applications. However, it is instructive to study this case as it ultimately yields the best pinning configurations for given external parameters. Consider the exemplary situation of a fixed magnetic field applied along the $z$-axis (or $c$-axis in HTSs) and current flowing along the $x$ direction. Na\"{i}vely, the optimal pinning landscape should mimic the vortex configuration for zero applied current at the given field, namely the Abrikosov vortex lattice. Hence, the pinscape should be a hexagonal array of columnar defects, with each column trapping a single vortex. However, the evolutionary approach yields an even better pinscape: a periodic array of planar pinning defects (`walls') that are aligned with the current and field direction (here parallel to the $xz$-plane).

In the simulation, we apply a constant external magnetic field $B = 0.1\Hct$ at low temperatures, corresponding to nearly zero noise (reduced temperature $\Tf = 10^{-5}$; see Sec.~\ref{sec:tdgl} for details). Inclusions are modeled by a non-superconducting material with zero critical temperature, $\Tci$, resulting in a suppressed order parameter, $\psi(\mathbf{r})$, inside the defects. Here $\Hct$ is the upper critical field at given temperature.

\begin{figure*}
	\begin{subfigure}[t]{0\linewidth} \phantomcaption \label{fig:planar_psi} \end{subfigure}%
	\begin{subfigure}[t]{\linewidth} \includegraphics[width=15cm]{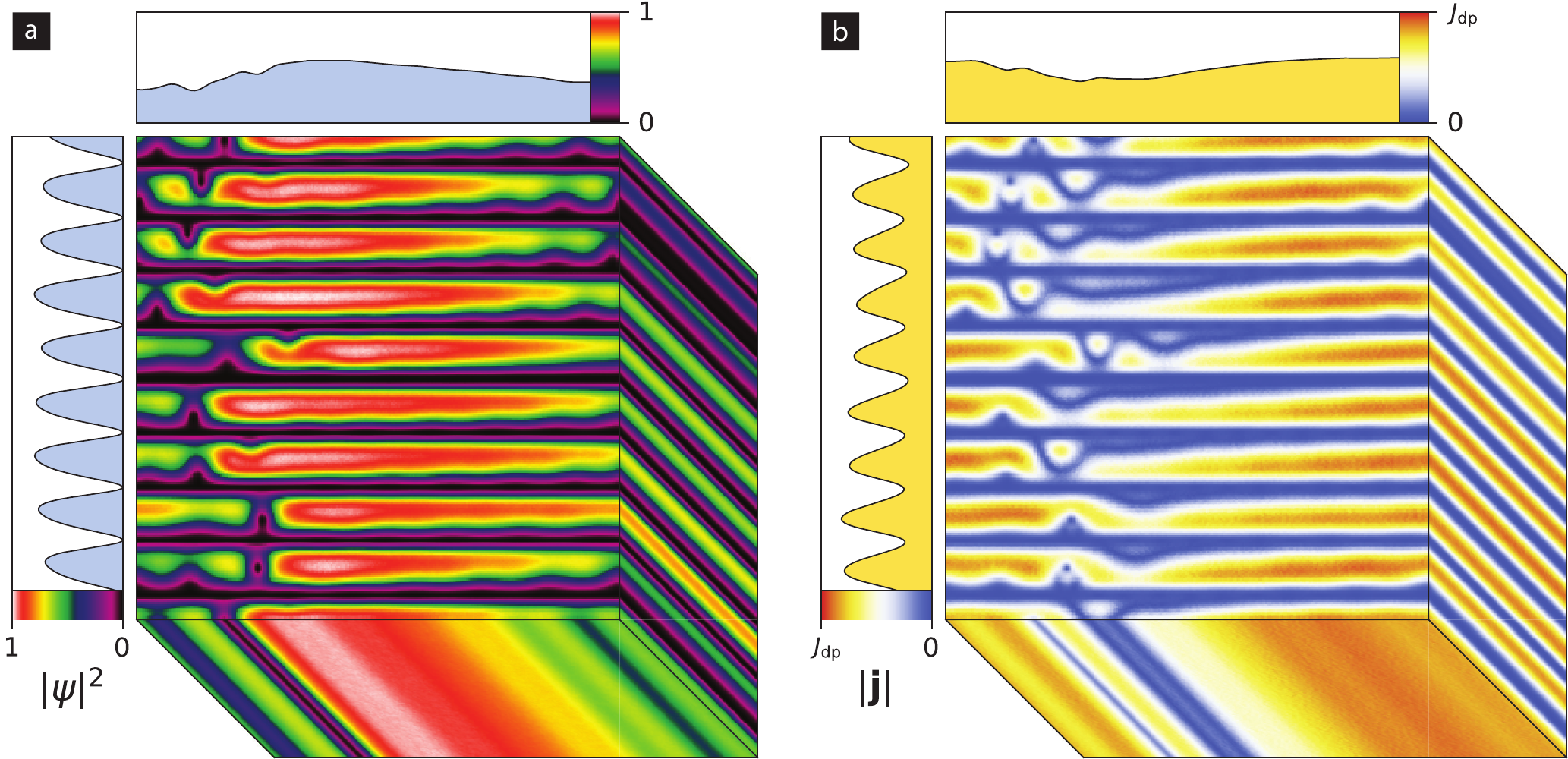} \phantomcaption \label{fig:planar_j} \end{subfigure}
	\caption{ \label{fig:planar_psi_j}
		Snapshot of \subref{fig:planar_psi} the squared order parameter amplitude $|\psi(\mathbf{r})|^{2}$ 
		and \subref{fig:planar_j} supercurrent amplitude $|\mathbf{j}_s(\mathbf{r})|^{2}$ for the periodic 
		planar defect configuration in the dynamic/dissipative regime with applied current slightly larger 
		than the critical current at magnetic field $B = 0.1\Hct$. The regions occupied by planar defects 
		have the suppressed order parameter [black horizontal planes in panel~\subref{fig:planar_psi}] and 
		zero supercurrent [blue planes in panel~\subref{fig:planar_j}]. The superconductor regions between 
		planar defects have mostly larger order parameter (shown in red and white) interrupted 
		by depinned vortices. Depinning events have distinctly collective behavior, i.e., all vortices depin 
		simultaneously in a certain region spanning through the system almost normal to the planar defects. 
		Due to geometrical constraints of the defects and strong vortex-vortex interaction, vortices 
		do not bend much, which is seen in the depth projection of the system. 
		The corresponding vortex dynamics is shown for magnetic fields 
		$B = 0.1\Hct$ (\uhref{https://youtu.be/TQpoO2piwlY}{movie clip}), 
		$0.2\Hct$ (\uhref{https://youtu.be/ZFBeFblNIS8}{movie clip}), and 
		$0.3\Hct$ (\uhref{https://youtu.be/BUc3yuylLHM}{movie clip}). 
		Vertical and horizontal insets show the order parameter [panel~\subref{fig:planar_psi}] and 
		supercurrent amplitude [panel~\subref{fig:planar_j}] averaged over other directions.
	}
\end{figure*}

\begin{figure}[tb] 
	\begin{subfigure}[t]{0\linewidth} \phantomcaption \label{fig:planar_iv} \end{subfigure}%
	\begin{subfigure}[t]{0\linewidth} \phantomcaption \label{fig:planar_psi_sc} \end{subfigure}%
	\begin{subfigure}[t]{\linewidth} \includegraphics[width=8.6cm]{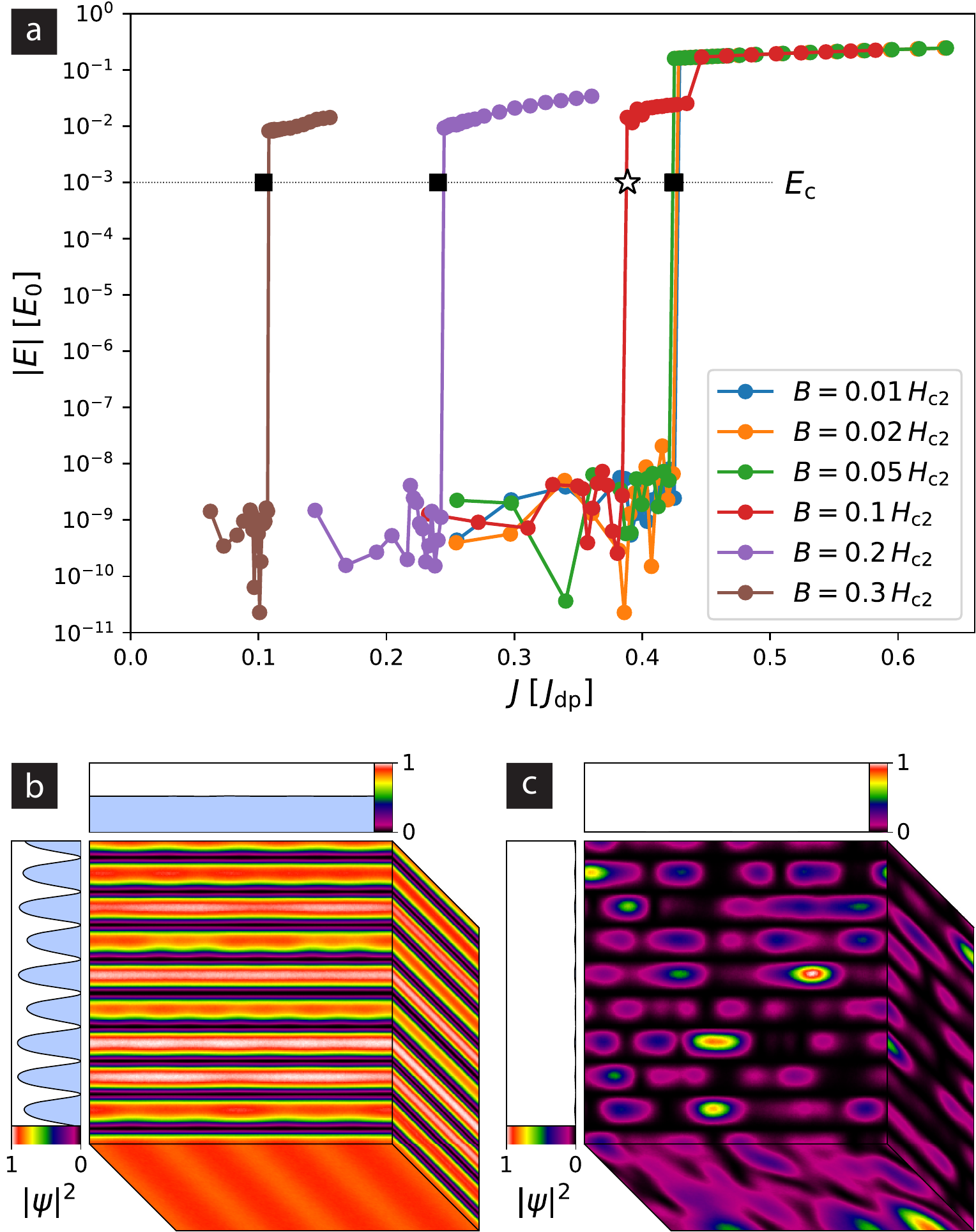} \phantomcaption \label{fig:planar_psi_fluct} \end{subfigure}
	\caption{ \label{fig:planar_iv_psi}
		\subfignum{(a)}~Current-voltage ($J$-$E$) curves for the planar defect pinning landscape, 
		which is optimal for field $B = 0.1\Hct$, in different applied magnetic fields. 
		In this regime each curve shows an extremely sharp drop (more than 6 orders of magnitude) 
		at the corresponding critical current (black square) determined by the finite voltage criteria 
		$\Ec = 10^{-3} E_0$ (dashed line). Voltage levels below $10^{-8} E_0$ cannot be resolved 
		due to numerical noise. 
		\subfignum{(b)}~Order parameter amplitude at $B = 0.05\Hct$ and current slightly below 
		the critical current, $J = 0.9999\Jc$ showing the superconducting state 
		(\uhref{https://youtu.be/cS_pSrae7so}{movie clip}). 
		\subfignum{(c)}~Order parameter amplitude at $B = 0.05\Hct$ and current slightly above 
		critical current, $J = 1.0001\Jc$ shows suppressed superconductivity 
		(\uhref{https://youtu.be/_g6BZooPJM8}{movie clip}) with localized superconducting regions.
	}
\end{figure}

The actual evolution process is illustrated in Fig.~\ref{fig:evolution}. As mentioned above, we start with the superconductor having no inclusions, shown in the left panel in Fig.~\ref{fig:evolution_snapshots}. In the {1\st}~generation we generate inclusions having random shapes and pick the configuration with highest critical current; this configuration consists of a single inclusion of almost spherical shape. In the {2\nd} and {3\rd}~generations, the spherical shape evolves into a new prolonged shape along the current. This prolongated ellipsoid has rather high pinning potential force for a few vortices and thus serves as `seed' inclusion for the rest of the generated inclusions. In the {4\th} generation, a copy of this inclusion is produced and placed at some distance from the original. In the next several generations, more and more copies of this inclusion are created to enhance the total pinning and, consequently, the critical current. Subsequent evolution moves these inclusions and modestly alters their shapes, which leads to a slight increase in the critical current. Note, that critical current rises faster in early generations; improvements in later generations require more mutations and lead to a smaller gain in critical current. The evolution terminates with the {37\th} generation and results in a set of almost equidistant planar defects oriented in the direction of applied current and having a thickness on the order of a coherence length, see Fig.~\ref{fig:evolution_final_snapshot}. The distance between planar defects roughly corresponds to positions of vortex rows in a perfect hexagonal lattice (blue circles) generated by the external magnetic field. The full evolution tree has $37$ generations and $6331$ pinning configurations, see Fig.~\ref{fig:evolution_path}. The best landscapes in each generation are numbered and have color ranges from blue with almost zero critical current to orange with maximal critical current $\Jc = 0.40\Jdp$, where $\Jdp$ is the depairing current. Each numbered configuration has at least $20$ successors: (i)~the successor with maximum critical current becomes the numbered seed for the next generation, (ii) all other successors are shown by small colored circles. These configurations have smaller critical currents than the seed and are discarded by our targeted selection. The final {37\th} configuration in the center has $2040$ mutations with smaller critical currents and thus considered as final. The vast majority ($90$\%) of these dead mutations lead to marginal decreases of the critical current (from $1$ to $15$\%, shown in green). Therefore, the determined configuration, shown in Fig.~\ref{fig:evolution_final_snapshot}, is rather stable with respect to mutation.

The used parameters produce a rather large critical current, $\Jc(T) = 0.40\Jdp(T)$ at almost zero temperature. For a larger noise level ($\Tf = 0.28$) corresponding to a temperature $T \sim 77$\,K, the critical current reduces to $\Jc(T) = 0.34\Jdp(T)$. Weaker metallic pinning centers with higher critical temperature, e.g., $\Tci = 2T - \Tcb$ ($\Tcb$ is the critical temperature in the bulk superconductor) can produce a maximal critical current of $\Jc = 0.31\Jdp$ at zero noise. In all these cases we can easily extrapolate a model for the optimal pinning configuration with only two parameters: the thickness of the planar defects and their separation.

Based on the optimal pinscape, we studied the critical dynamics close to $\Jc$. A snapshot of the order parameter and supercurrent density amplitude is presented in Fig.~\ref{fig:planar_psi_j}. Here, we set the field to $B = 0.1\Hct$ and applied a current $J$ slightly larger than the critical current $\Jc$, $J = 1.0001\Jc$. The depinning process defining the critical current occurs via a collective avalanche  across the sample in a narrow channel along the Lorentz force (normal to the planar defects). Single vortex motion never occurs; instead, if a vortex depins from one planar defect, vortices from the neighboring defects also depin to either free space for the vortex or fill its vacant position. This collective depinning effectively increases the pinning force of the system. The same collective behavior occurs for other types of pinning landscapes, which are optimized for highest possible critical currents, e.g., for ordered defects\cite{Sadovskyy:2017hex} or disordered nanorods extended along the direction of the applied magnetic field.\cite{Sadovskyy:2016jc} A similar but somewhat less pronounced effect was observed for randomly placed spherical particles.\cite{Koshelev:2016} Due to such system-spanning clusters of collectively depined vortices in pinscapes with very large critical currents, the dynamic transitions to the dissipative state tend to be more pronounced than for sub-optimal configurations showing single-vortex depinning. This abrupt transition can be also seen in the current-voltage ($J$-$E$) curves shown in Fig.~\ref{fig:planar_iv}. Here, the various $J$-$E$ curves are associated with different applied magnetic fields, $B$, for the same pinning landscape, which was optimized for maximum $\Jc$ at $B = 0.1\Hct$. Each curve displays a sharp transition with relative voltage drop of at least six orders of magnitude around its critical current shown by a star in the data for $B = 0.1\Hct$ and black squares for other $B$ values. Note, that for lower magnetic fields ($B \lesssim 0.05\Hct$), the superconducting state with pinned vortices at $J < \Jc$ [Fig.~\ref{fig:planar_psi_sc}] transits directly to a dissipative state consisting of only localized superconducting regions for $J > \Jc$ [Fig.~\ref{fig:planar_psi_fluct}], which cannot pass a supercurrent through the system, hence bypassing the dissipative superconducting state.

Such sharp transitions allow, in particular, the use of a finite-voltage criterion to determine the critical current with rather high threshold electric field, $\Ec$, as shown by the horizontal dashed line in Fig.~\ref{fig:planar_iv}. This threshold field can be many orders of magnitude larger than the $1\,\mu$V criterion typically used in experiments, which dramatically reduces the computation time for a single $\Jc$ estimation.

Our targeted evolution derived pinning landscape with critical current $\Jc = 0.40\Jdp$ can be compared to other typical pinscapes with potentially high critical currents at the same magnetic field $B = 0.1\Hct$ and low thermal noise: 
\begin{itemize}
\item[(i)] Randomly placed spherical defects with optimal diameter and concentration have a maximum possible critical current $\Jc = 0.061\Jdp$.\cite{Koshelev:2016}
\item[(ii)] Field aligned randomly placed columnar inclusions with best diameter and concentration lead to $\Jc = 0.091\Jdp$.\cite{Kimmel:2017a}
\item[(iii)] Hexagonally ordered field-aligned columnar defects with optimal size and concentration, generate a significantly larger critical current $\Jc=0.32\Jdp$, but still smaller than for planar defects.
\end{itemize}
Next, we compare the properties of hexagonally ordered columnar with that of arrays of planar defects derived from the genetic approach.

\begin{figure*}
	\begin{subfigure}[t]{0\linewidth} \phantomcaption \label{fig:Jc_B_planar} \end{subfigure}%
	\begin{subfigure}[t]{0\linewidth} \phantomcaption \label{fig:Jc_B_hex} \end{subfigure}%
	\begin{subfigure}[t]{0\linewidth} \phantomcaption \label{fig:Jc_Bm_planar} \end{subfigure}%
	\begin{subfigure}[t]{\linewidth} \includegraphics[width=16.1cm]{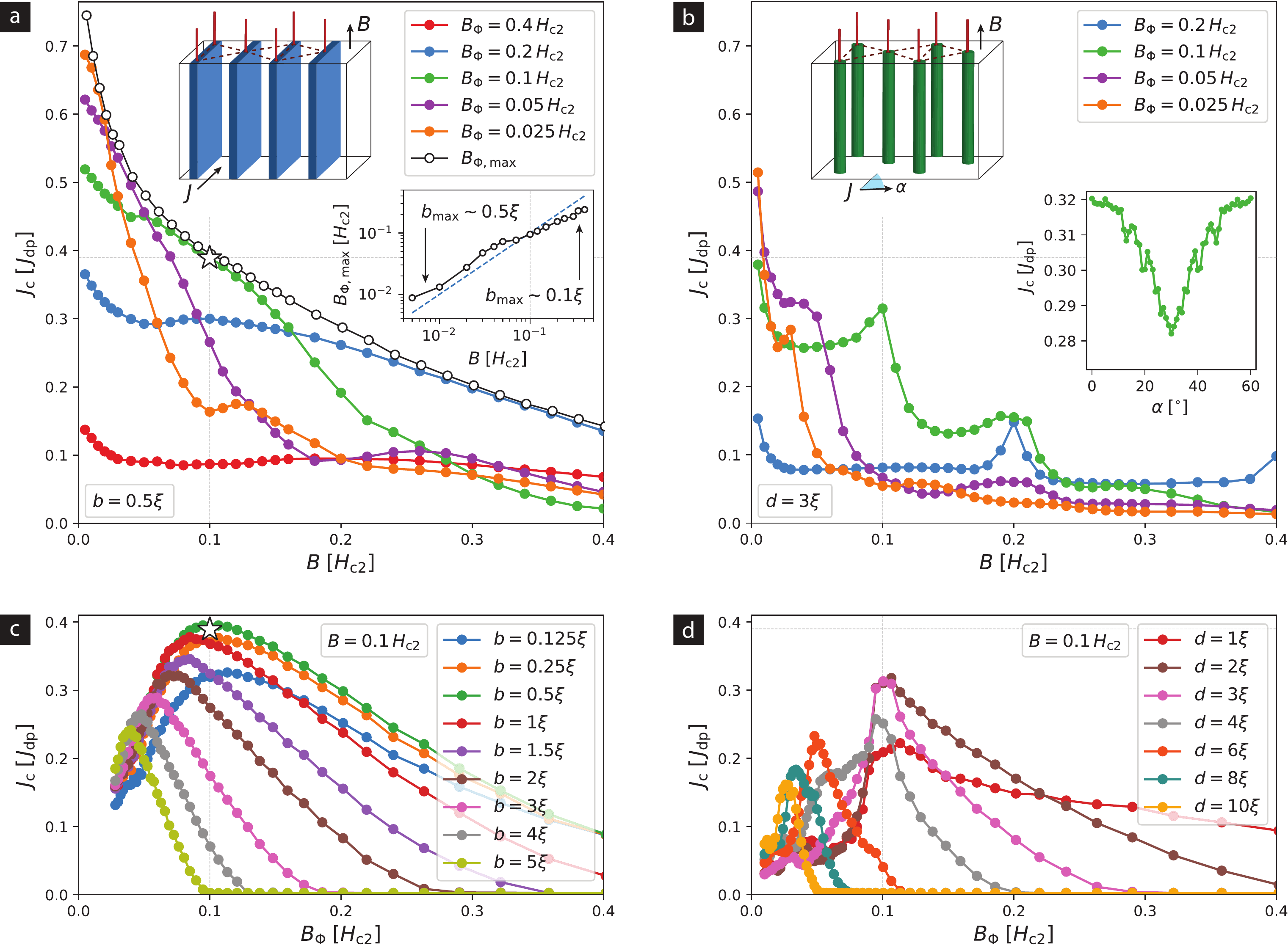} \phantomcaption \label{fig:Jc_Bm_hex} \end{subfigure}
	\caption{ \label{fig:Jc_B_Bm}
		Critical current as a function of magnetic field, $\Jc(B)$, for different landscapes 
		and matching fields $\Bm$. 
		\subref{fig:Jc_B_planar}~Planar defects with fixed thickness of $b = 0.5\xi$ and range 
		of $\Bm$ from 0.025 to $0.4\Hct$. The star shows the targeted evolution result 
		for $B = 0.1\Hct$. The envelope curve (black line with open circles) shows 
		the maximal possible $\Jcmax(B)$ at a given field $B$. 
		The inset shows the corresponding optimal matching field $\Bmmax$, 
		i.e., the distance needed to achieve this maximum. 
		\subref{fig:Jc_B_hex}~Hexagonal pattern of columnar defects with diameter $d = 3\xi$. 
		The inset shows $\Jc$ as a function of hexagonal lattice rotation angle $\alpha$ 
		with respect to the applied current. It is $\pi$/3-periodic and maximal if the current 
		is aligned with the lattice axes. 
		\subref{fig:Jc_Bm_planar}.~Critical current as a function of matching field for planar defects 
		in applied field $B = 0.1\Hct$ for different wall thicknesses, $b$. 
		\subref{fig:Jc_Bm_hex}.~The same for columnar defects ordered in a hexagonal pattern 
		for different diameters, $d$, of the columns.
	}
\end{figure*}

\section{Planar vs. columnar defects} \label{sec:columns}

A systematic comparison of a hexagonal lattice of columnar defects to the extrapolated model of a periodic array of planar defects requires comparable parameters. The natural parameters for columnar defects are the matching field~$\Bm$ (the hypothetic magnetic field producing an Abrikosov vortex lattice with the same density as the lattice of columns) and their diameter, $d$. For arrays of planar defects, one can use the same matching field $\Bm$ and place the defects along one of the main axes (parallel to the current) of the hexagonal lattice (and along the field), see insets in Fig.~\ref{fig:Jc_B_planar} and \ref{fig:Jc_B_hex}. The distance between the planar defects is then $h = 3^{1/4}\xi(\pi\Hct/\Bm)^{1/2}$. The second parameter is the thickness of the planar defects. In both cases the maximum critical current is reached when $B = \Bm$, see Figs.~\ref{fig:Jc_Bm_planar} and \ref{fig:Jc_Bm_hex}. A main difference is that the planar-array is more robust against changes in $\Bm$ than the discrete columnar defects structure, i.e., small changes in $\Bm$ [or $h$] result in very small changes in the optimal critical current. 

Figure~\ref{fig:Jc_B_planar} shows the $\Jc(B)$ dependence for the planar array with fixed thickness, $b = 0.5\xi$, and different $\Bm$ ranging from $0.025\Hct$ to $0.4\Hct$. All curves display a relatively `smooth' behavior. The most representative is the green curve simulated for $\Bm = 0.1\Hct$. The open star corresponds to the critical current associated with the pinning landscape shown in Fig.~\ref{fig:evolution_final_snapshot} obtained by targeted evolution for $B = 0.1\Hct$. The envelope curve (black line with circles) shows $\Jcmax(B)$, the critical current for optimized landscapes for each fixed $B$ with optimal wall thickness $b(B)$ and matching field $\Bmmax(B)$. The deviation of $\Bmmax(B)$ from a simple linear dependence $\Bm = B$ (see inset) is due to different $b(B)$, ranging from $\sim 0.5\xi$ at low fields to $\sim 0.1\xi$ at higher fields.

Figure~\ref{fig:Jc_B_hex} shows the $\Jc(B)$ dependence for hexagonal patterned columnar defects with fixed diameter $d = 3\xi$ for different $\Bm$ from $0.025\Hct$ to $0.2\Hct$. The green curve shows a peak at the first matching field with $\Jc = 0.32\Jdp$, which coincides with the maximal $\Jc$ of the hexagonal lattice at $B = 0.1\Hct$. A rotation of the hexagonal pattern can reduce this value (it is maximum if a main axis of the lattice is aligned with the current), see angular dependence in the inset.

Ordered planar defects optimized for some certain magnetic field, $B$, have higher critical current than ordered columnar defects optimized for the same field. In addition, in the case of planar-defects pinscape, the critical current, $\Jc(B')$, is a monotonic function of magnetic field $B'$ below $B$ at which the pinscape was optimized, i.e. $\Jc(B') \geqslant \Jc(B)$ for $B' \leqslant B$. Therefore, the optimized critical current of planar defects cannot drop for smaller magnetic field, which does not hold for ordered columnar defects with non-monotonic $\Jc(B')$.

Figure~\ref{fig:Jc_Bm_planar} depicts the $\Jc$($\Bm$)-dependence for arrays of planar defects with different wall thickness $b$ at fixed applied field $B = 0.1\Hct$. This sampling shows a single robust optimum near $\Bm = 0.1\Hct$ and $b = 0.5\xi$. A similar sampling for the hexagonal columnar defect pattern presented in Fig.~\ref{fig:Jc_Bm_hex} shows significantly sharper peaks in the vicinity of the matching field, resulting in less robust behavior against small changes of the parameters. $\Jc$-samplings for other parameters are shown in Figs.~\ref{fig:Jc_l_b_hex_cols}--\ref{fig:Jc_a_b_77K} in Sec.~\ref{sec:sampling}.

\begin{figure*}
	\begin{subfigure}[t]{0\linewidth} \phantomcaption \label{fig:pre_existing_planar} \end{subfigure}%
	\begin{subfigure}[t]{0\linewidth} \phantomcaption \label{fig:pre_existing_ellipses} \end{subfigure}%
	\begin{subfigure}[t]{\linewidth} \includegraphics[width=16.8cm]{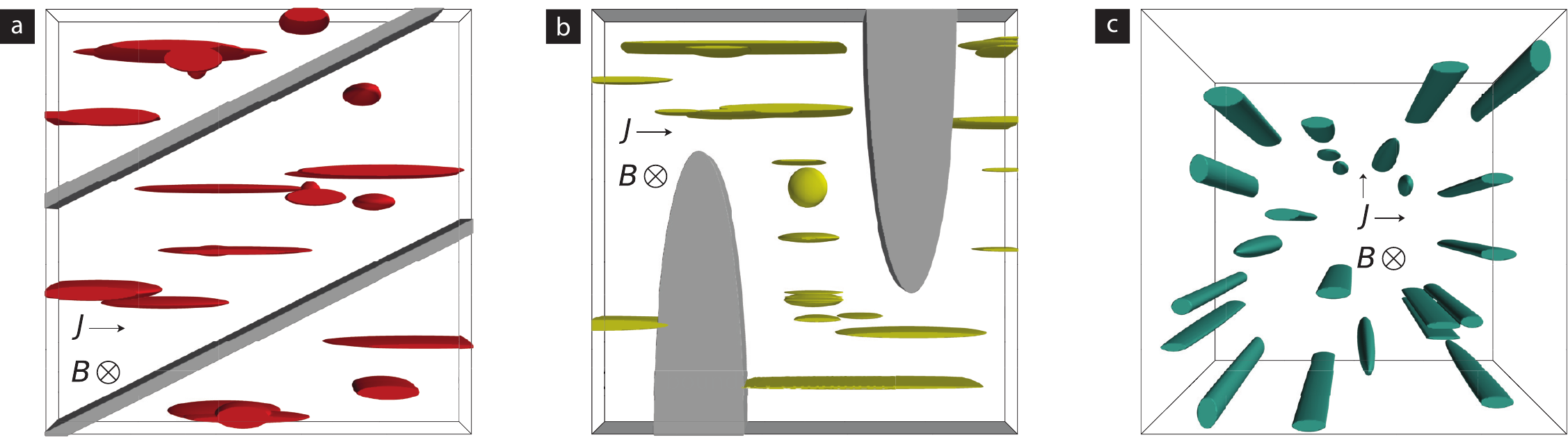} \phantomcaption \label{fig:undirected_current} \end{subfigure}
	\caption{ \label{fig:applications}
		Evolution of pinning landscapes for predefined environments. 
		\subref{fig:pre_existing_planar}~Current is applied from left to right and magnetic field is fixed 
		at $B = 0.1\Hct$ perpendicular to the figure plane as in Fig.~\ref{fig:evolution}. 
		The difference is in the pre-existing pinscapes containing tilted planar defects 
		shown in grey. These plates redirect the supercurrent flow (boundary conditions 
		are periodic in the figure plane) and make the optimal pinscape shown 
		in Fig.~\ref{fig:evolution} inefficient. The evolutionary approach generates 
		smaller `planar' defects (or flat cylinders) along the current between 
		the pre-existing inclusions. 
		\subref{fig:pre_existing_ellipses}~In this scenario, two flattened half-cylinders block the current 
		with open (no-current) boundary conditions at top and bottom surfaces. 
		Generated inclusions are more cylindrical between inclusions 
		to avoid blocking supercurrents. 
		\subref{fig:undirected_current}~In this scenario, the current can be applied in left-right and 
		bottom-up directions and the largest critical current is defined by the minimum 
		critical current in either direction. The pinscape evolves to hyperuniformly 
		placed columnar defects.
	}
\end{figure*}

To understand why planar defects give a larger critical current compared to columns with circular cross section, let us consider an isolated cylindrical defect of elliptical cross section with main axes $a$ and $b$. We apply an average current density $J$ along $a$ and a magnetic field along the cylinder. The local current density at the depinning point (at the extremal points of the defect boundary along $b$) is $\Jl = J(a+b)/a$; for columnar defects with circular cross section $\Jl = 2J$, while for planar defects this value is twice smaller, $\Jl = J$. It means that the depinning force is expected to be twice larger for a planar defect than for a columnar defect at the same average current density. In the case of an ordered pattern of defects the situation is more complex. Figures~\ref{fig:Jc_l_b_planar}--\ref{fig:Jc_l_d} in Sec.~\ref{sec:sampling} show critical current for a hexagonal lattice made of columnar defects with diameters $a$ and $b$ such that the $a/b$ ratio ranges from planar defects ($a \to \infty$) to cylindrical columns ($a = b$). This systematic sampling shows that critical current tends to increase with the ratio $a/b$ for the same defect area $\pi ab/4$ for strong pinning defects. For example, columnar defects with $a = 4\xi$ and $b = \xi$ produce $\sim 13\%$ larger critical current comparing to cylindrical defects with $a = b = 2\xi$. Further increase of the ratio $a/b$ converts hexagonal array of columnar defects to the array of the planar defects, however the optimal thickness of each planar defect is determined by its material properties.

\section{Applications of targeted evolution} \label{sec:applications}

A recent report on doubling the critical current of commercial HTS wire by additional particle irradiation\cite{Jia:2013} highlights the importance and advantages of a post-synthesis approach to enhance the critical current, while leaving the wire synthesis process untouched. Our targeted evolution approach can also be applied to systems with pre-existing defects. Figure~\ref{fig:applications} demonstrates results of targeted evolution in different environments, defined by either pre-existing pinscapes or different external parameters. In Figs.~\ref{fig:pre_existing_planar} and \ref{fig:pre_existing_ellipses}, we apply the evolutionary algorithm to pinscape with fixed pre-existing defects. These defects partially block the left-to-right current flow and, thus, dramatically change the result of the targeted evolution described above. Mainly, the evolution leaves some defect-free regions in the superconducting matrix to allow for a supercurrent path. In the case of pre-existing tilted walls in Fig.~\ref{fig:pre_existing_planar}, the total current $\Ic = \Jc w t$ through the system was increased by evolution from $\Ic = 56 \Jdp \xi^2$ ($\Jc = 0.11\Jdp$), to $\Ic = 147 \Jdp \xi^2$ ($\Jc = 0.29\Jdp$) in applied field $B = 0.1\Hct$, where $w = 64\xi$ and $t = 8\xi$ are the system's width and thickness, respectively. In the case of pre-existing two half-ellipses shown in Fig.~\ref{fig:pre_existing_ellipses}, the critical current rises, from $\Ic = 35 \Jdp \xi^2$ ($\Jc = 0.068\Jdp$) to $\Ic = 104 \Jdp \xi^2$ ($\Jc = 0.20\Jdp$) upon evolution of added defects.

In Fig.~\ref{fig:undirected_current} we apply the current both in the horizontal and vertical directions and consider the fitness function $\Jcu = \mathrm{min}\{\Jcright, \Jcup\}$, where $\Jcright$ is left-to-right $\Jc$ and $\Jcup$ is bottom-to-up $\Jc$, rather than only $\Jcright$ as before. $\Jcu$ approximately models arbitrary directions of applied currents. The resulting pinscape consists of columnar defects along the magnetic field arranged in a hyperuniform `pattern'.\cite{Sadovskyy:2017hex, Thien:2017} The corresponding critical current density, $\Jc = 0.27\Jdp$, is 5\% less than the $\Jc$ for a hexagonal lattice oriented in the `wrong' way [rotated $\pi/6$ from the main axes, see the angular dependence in the inset of Fig.~\ref{fig:Jc_B_hex}].

In all the simulations above, we intentionally did not limit the size, shape, or placement of the mutated defects. However, it is possible to limit the defect morphology to mimic the limitations of, e.g., (i)~oxygen irradiation, which creates point defects of certain size distribution, (ii)~heavy ion irradiation, which creates tracks/columns, or (iii)~chemically grown defects, which cannot overlap.

\section{Discussion and conclusions}

In this paper, we introduced an evolutionary approach for the optimization of pinscapes in type-II superconductors. This approach utilizes the idea of targeted selection inspired by biological natural selection. We demonstrated that it can be applied to enhance the current carrying capacity of superconductors in a magnetic field.

We discovered that certain patterns of defects composed of metallic inclusions can maximize the critical current up to 40\% of $\Jdp$ for fixed direction of the current perpendicular to the magnetic field at 10\% of $\Hct$. We numerically demonstrated that that no other mixture of different defect shapes can reach this level of $\Jc$. The discussed pining structure may arise in niobium titanium wires, in which a sequence of heating/drawing steps result in a microstructure composed of nanometer-scale metallic and almost parallel $\alpha$-titanium lamellae embedded in the niobium titanium matrix.\cite{Lee:2003} Furthermore, the layered structure of cuprate HTSs give rise to intrinsic pinning of similar nature.

In contrast to conventional optimization techniques such as coordinate descent, where one varies only a few global parameters characterizing the entire sample (e.g., size and concentration of defects), our targeted evolution approach allows us to vary each defect individually without any \textit{a priori} assumptions about the defects configuration. This flexibility outweighs its higher computational cost. The considered optimization problem has basically infinite degrees of freedom, prompting one to ask why the evolution method convergences relatively quickly. One reason is that there are a lot of configurations with critical currents quite close to the maximum possible one, which are in practice, indistinguishable from each other. The evolutionary approach just allows us to find one such configuration. Typically, larger regions of near-optimum configurations correspond to a broader maximum of $\Jc$ as a function of a set of appropriate parameters, e.g., the system in Fig.~\ref{fig:Jc_Bm_planar} evolutionarily adapts faster than the system in Fig.~\ref{fig:Jc_Bm_hex}.

We also demonstrated the enhancement of the critical current for two cases of pre-existing defects, found in commercial HTSs. Our approach provides a computer assisted route to rational enhancement of the critical current in applied superconductors. It can be used to define a post-synthesis optimization step for existing state-of-the-art HTS wires for high-field magnet applications by modeling the actual geometry of the wire within the magnet and taking into account external magnetic field distributions and self-fields. This can be done by coupling TDGL equation with Maxwell equations and initiate the simulation with a pre-existing defect distribution in the wire.

Finally, we note that the described evolutionary algorithm is a local method and thus can easily get stuck in a local maximum. An analogue in biological evolution is the extreme detour of a giraffe's recurrent laryngeal nerves,\cite{mammal_anatomy:2010} which became `trapped' under the aortic arch in the thorax. However, in contrast to natural selection, targeted evolution can be performed multiple times. Namely, a comparison of the resultant pinscapes and corresponding critical current values allows us to estimate how close they are to the best possible pinscape, making targeted evolution \textit{global}. Moreover, by finding different near-maximum points, it is possible to understand which parameters are important for large critical current and which ones are not. An experimental analogue in organic systems is the process of \textit{in vitro} selection.\cite{Jijakli:2016} A particular example is the selection of RNA molecules being able to bind to specific ligands:\cite{Ellington:1990} it was shown that `evolved' molecules bind stronger than those of the first generation and an \textit{a-priory} `guess' of the best binding RNA sequence would not have been possible.

In conclusion, our methodology of utilizing targeted evolutionary concepts to improve the intrinsic properties of condensed matter systems is a promising path towards the design of tailored functional materials. It can be applied to a large variety of different physical systems, and has demonstrated its usefulness in the enhancement of superconducting critical currents. Furthermore, its ability to take existing environments into account, allows for optimization by post processing.

\paragraph*{Acknowledgements.} We thank L.\,Civale, R.\,Willa, and I.\,S.\,Aranson for numerous useful comments. I.\,A.\,S, A.\,E.\,K., and A.\,G. were supported by the Scientific Discovery through Advanced Computing (SciDAC) program funded by U.\,S. Department of Energy (DOE), Advanced Scientific Computing Research. U.\,W. and W.-K.\,K. were supported by the Center for Emergent Superconductivity, an Energy Frontier Research Center, funded by DOE, Office of Basic Energy Sciences. Simulations were performed at the Oak Ridge Leadership Computing Facility (DOE contract DE-AC05-00OR22725), the Argonne Leadership Computing Facility (DOE contract DE-AC02-06CH11357), and the Computing Facility at Northern Illinois University. 
Other simulations of superconducting vortices can be found at \uhref{https://www.youtube.com/channel/UCjdQ4Ruhxma5pkxGrFxw3CA}{YouTube channel} and \uhref{https://goo.gl/ewcct7}{OSCon website}.

\bibliography{evolution}
\balance

\onecolumngrid
\appendix
\clearpage 

\renewcommand\appendixname{}
\renewcommand{\thesection}{S\arabic{section}}
\renewcommand\thefigure{S\arabic{figure}}

\section{Time-dependent Ginzburg-Landau model} \label{sec:tdgl}

To model vortex dynamics in the superconductor, we solve the dimensionless TDGL equation, 
\begin{equation} \label{eq:tdgl}
	(\partial_t + \i\mu) \psi 
	= \varepsilon(\r) \psi - |\psi|^2 \psi
	+ (\nabla + \i\A)^2\psi + \zeta(\r, t)
\end{equation}
for the complex order parameter $\psi(\r,t)$ in the infinite-$\lambda$ limit. We evolve a system of size $64\xi \times 64\xi \times 8\xi$ with grid spacing of 0.5$\xi$ and quasi-periodic boundary conditions in each direction. 
Here, $\mu = \mu(\r)$ is the electric scalar potential, $\A$ the vector potential associated with the external magnetic field $\B = \nabla \times \A$, and $\zeta(\r,t)$ is the temperature-dependent $\delta$-correlated Langevin term. The unit of length is the superconducting coherence length $\xi = \xi(T)$ at a given temperature $T$, the unit of magnetic field is the upper critical field, $\Hct = \Hct(T) = \hbar c/2e\xi^2$, the unit of the current density,
\begin{equation} \label{eq:j}
	\J 
	= \frac{3^{3/2}}{2} 
	\mathrm{Im} \Bigl\{ \psi^*(\nabla - \i\A)^2\psi 
	- \nabla\mu \Bigr\}
\end{equation}
is the depairing current, $\Jdp = \Jdp(T)$, and the unit of electric field is $E_0 = (3^{3/2}/2) \, \Jdp/\sigma$, where $\sigma$ is the normal-state conductance.

For the sampling shown in Figs.~\ref{fig:Jc_B_Bm} and \ref{fig:Jc_l_b_planar}--\ref{fig:Jc_a_b_77K} as well as for the conventional optimization in Fig.~\ref{fig:Jc_B_planar} we use a system of size $128\xi \times 128\xi \times 128\xi$. In the scenario shown in Fig.~\ref{fig:undirected_current} we used a $32\xi \times 32\xi \times 32\xi$ simulation box. 

In order to determine the critical current density, $\Jc$, we utilize a finite-electrical-field criterion. Technically, we adjust the applied external current to reach certain electrical-field level across the system. By targeting a small threshold electrical-field level $\Ec = 10^{-3}E_0$ and averaging over steady state long enough we obtain a critical current, see Refs.~\onlinecite{Sadovskyy:2015gl, Sadovskyy:2016real} for details.

\begin{figure}[h]
	\centering \includegraphics[width=4cm]{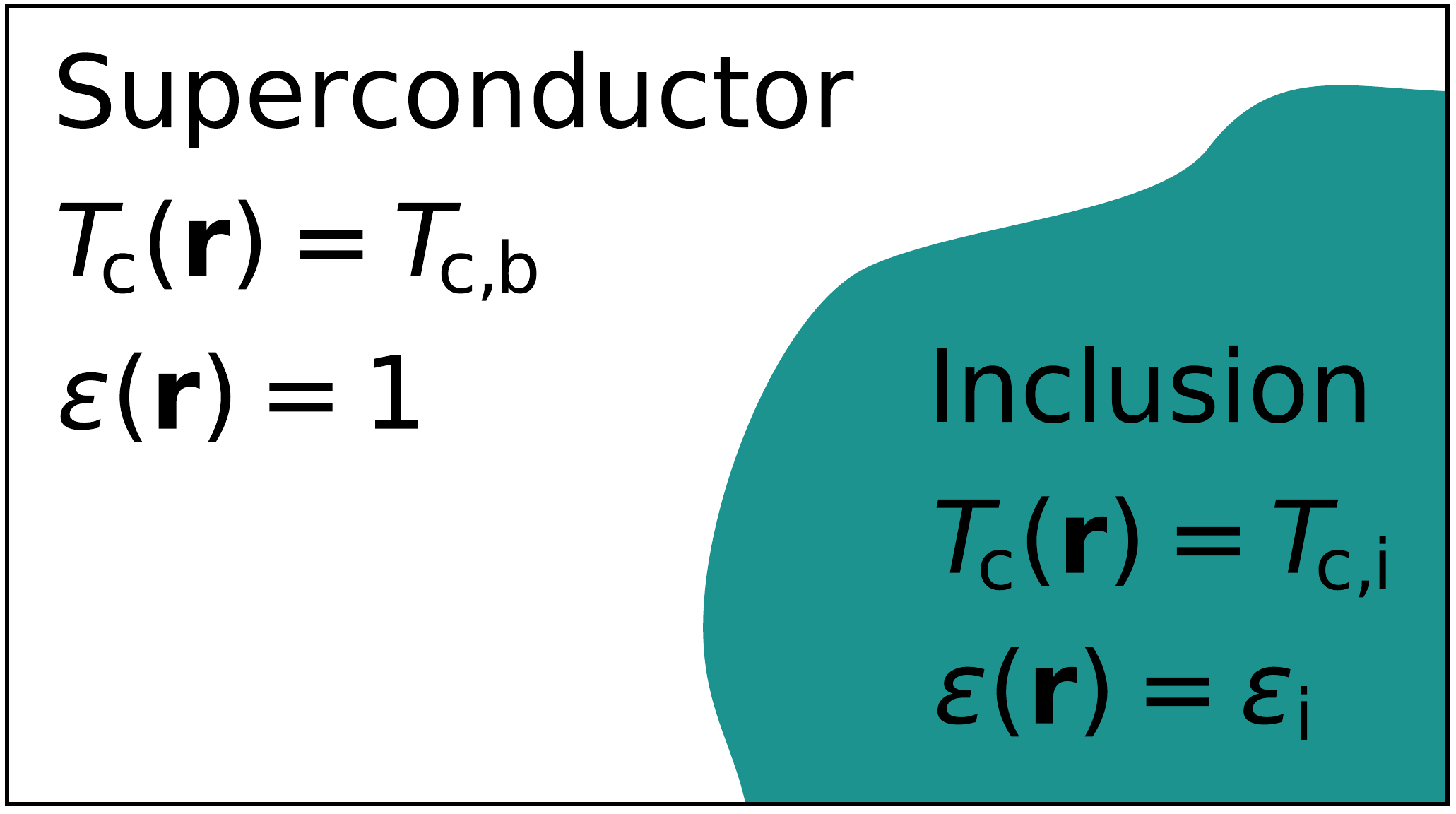}
	\caption{ \label{fig:Tc_map}
		Sketch of a critical temperature map.
	}
\end{figure}

In simulations we parametrize real temperature, $T$, and critical temperature map, $\Tc(\r)$ using two dimensionless quantities: linear term coefficient, 
\begin{equation} \label{eq:epsilon}
	\varepsilon(\r)
	= [\Tc(\r) - T] / [\Tcb - T],
\end{equation}
shown in Fig.~\ref{fig:Tc_map} and noise-level coefficient, $\Tf \propto T$, in Langevin term, $\zeta(\r,t)$.

We model non-superconducting inclusions with a reduced critical temperature, $\Tci$, inside each defect ellipsoid, resulting in a suppressed Ginzburg-Landau order parameter inside these inclusions. Typically, we use almost zero $\Tci$ corresponding to the suppressed order parameter in the entire inclusion region. Technically, we set a very negative coefficient before linear term, $\ei = (\Tci - T)/(\Tcb - T) = -30$, where we set $1 - T/\Tcb = 1/31$ is the critical temperature in the bulk superconductor and $T = \Tcb$ is a system temperature. In Figs.~\ref{fig:Jc_l_b_planar}--\ref{fig:Jc_a_b_77K}, we compare this situation to weaker pinning centers having $\ei = -1$ corresponding to $\Tci = 2T - \Tcb$.

The temperature-induced noise is modelled by the additive Langevin term, $\zeta(\r,t)$, in the TDGL equation with correlator $\langle \zeta^*(\r,t) \zeta(\r',t') \rangle = \Tf \, \delta(\r-\r') \delta(t-t')$, see Ref.~\onlinecite{Sadovskyy:2015gl} for details. In Figs.~\ref{fig:Jc_l_d}--\ref{fig:Jc_a_b_77K}, we compare the low-temperature regime, $T = 0$\,K, modeled by low Langevin-term coefficient $\Tf = 10^{-5}$ to the high-temperature regime, $T = 77$\,K, having high Langevin coefficient $\Tf = 0.28$.

\section{Details of numerical simulations}

We have implemented the time-dependent Ginzburg-Landau (TDGL) solver on General Purpose Graphics Processing Units (GP GPUs) using the CUDA framework and used Python for the evolutionary algorithm and job control on specialized computational clusters. The results of the evolution of different types were obtained on Titan, a Cray XK7 supercomputer at Oak Ridge Leadership Computing Facility running NVIDIA Kepler GPUs. We parallelized the computations running 16--256 pinning landscapes in one generation. We also used the high-performance GPU clusters GAEA at Northern Illinois University and Cooley at the Argonne Leadership Computing Facility for extrapolation, analysis, and visualizing the results.

\clearpage
\section{Sampling of critical current} \label{sec:sampling}

\begin{figure}[ht]
	\centering \includegraphics[width=17cm]{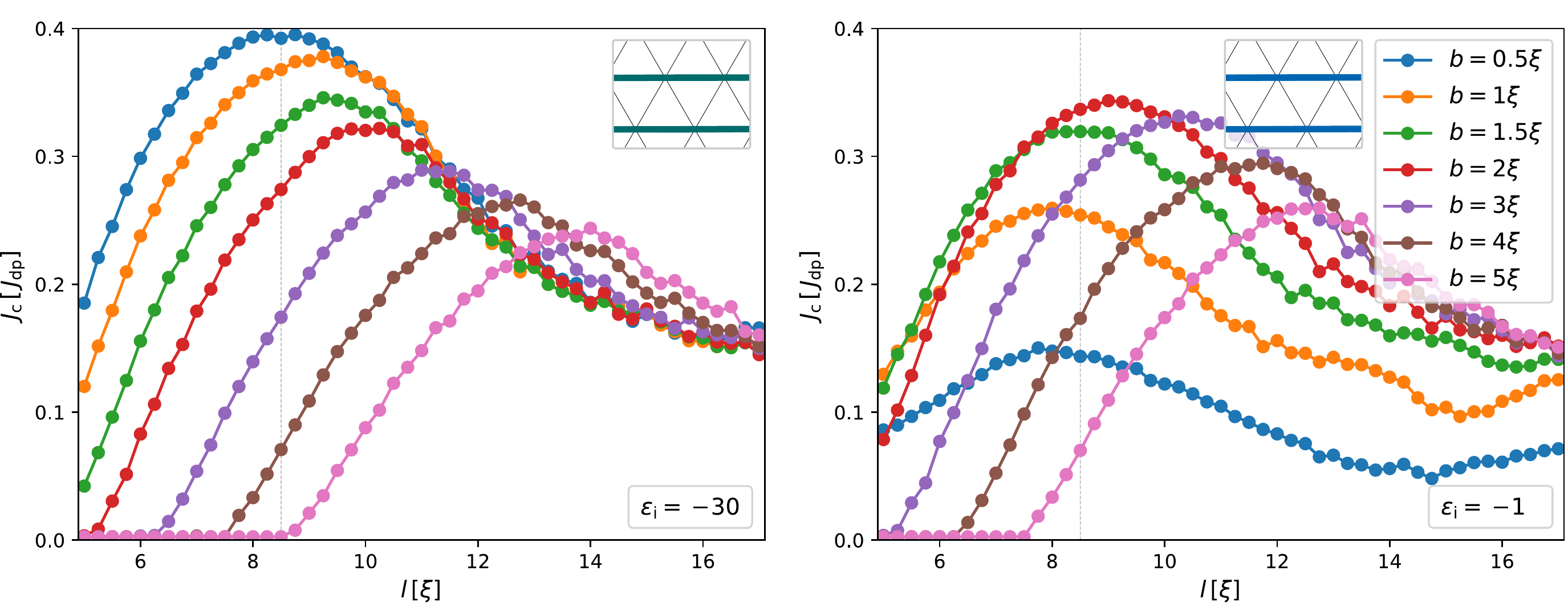}
	\caption{ \label{fig:Jc_l_b_planar}
		Critical current $\Jc$ for walls in $xz$ plane as a function of associated lattice 
		constant $l$ (proportional to the distance between planar defects $l = 2d/3^{1/2}$) 
		for different wall thickness $b$ in the cases of strong $\ei = -30$ (left) and 
		weak $\ei = -1$ (right) pinners. 
		A small Langevin noise coefficient is used ($\Tf = 10^{-5}$) corresponding to 
		near-zero temperatures.
	}
\end{figure}

\begin{figure}[ht]
	\centering \includegraphics[width=17cm]{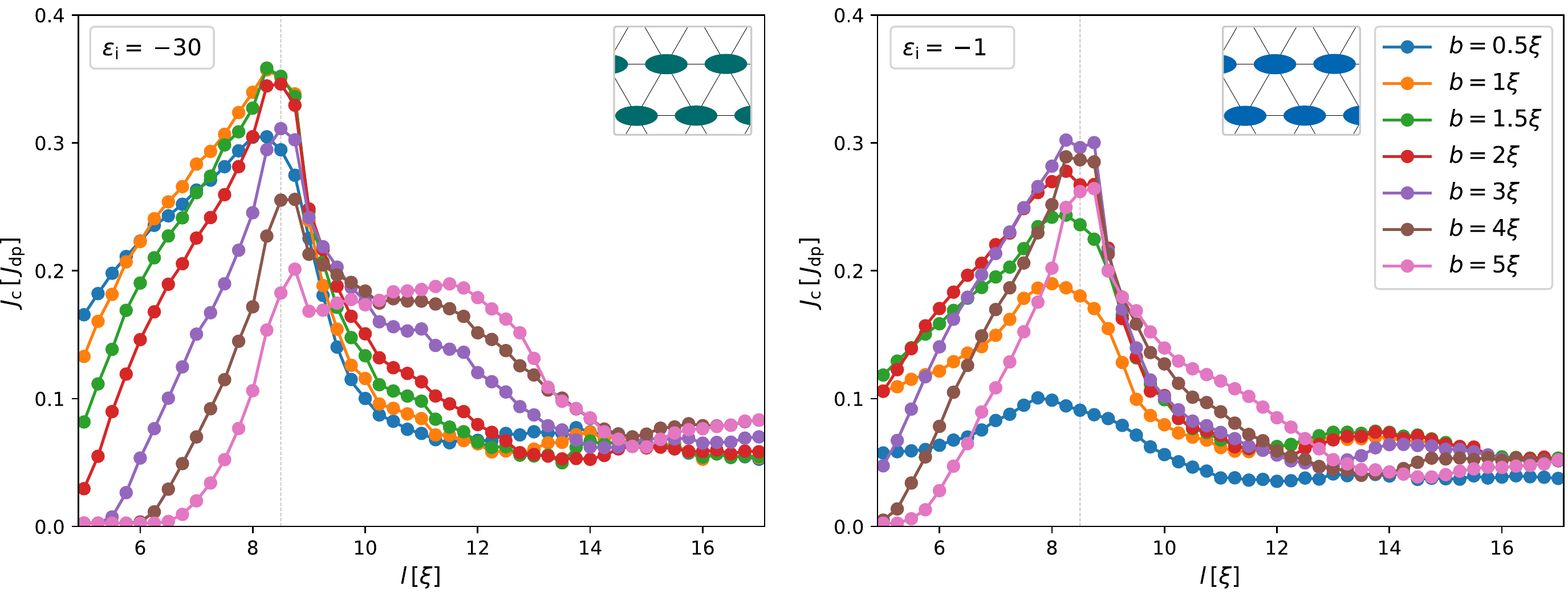}
	\caption{ \label{fig:Jc_l_b_hex_cols}
		Critical current $\Jc$ for hexagonal lattice of columnar defects with elliptical 
		cross section (diameter $a$ is in $x$ direction and $b$ is in $y$ direction) 
		as a function of lattice constant $l$ for a = $4\xi$ and different $b$ in the cases 
		of strong $\ei = -30$ (left) and weak $\ei = -1$ (right) pinners. 
		A small Langevin noise coefficient is used corresponding to near-zero temperatures.
	}
\end{figure}

\begin{figure}[ht]
	\centering \includegraphics[width=17cm]{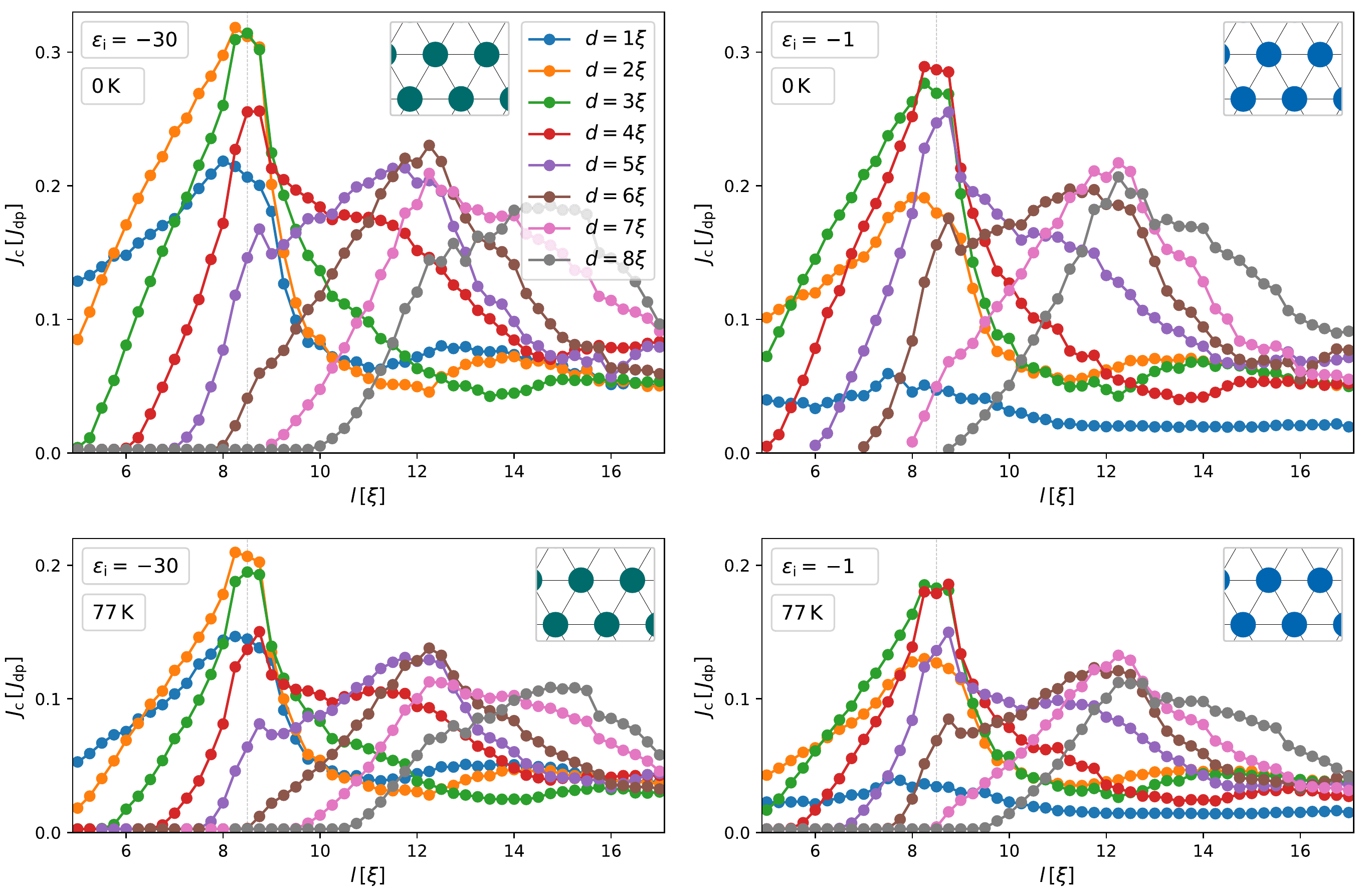}
	\caption{ \label{fig:Jc_l_d}
		Critical current $\Jc$ for hexagonal lattice of columnar defects with circular 
		cross section of diameter $d$ as a function of lattice constant $l$ 
		for different $d$ in the cases of strong $\ei = -30$ (left) 
		and weak $\ei = -1$ (right) pinners. Both are shown for two different temperatures: 
		small Langevin noise coefficient ($\Tf = 10^{-5}$) corresponding to near-zero 
		temperatures (top row) and larger noise coefficient ($\Tf = 0.28$) 
		corresponding to 77\,K (bottom row).
	}
\end{figure}

\begin{figure}[ht]
	\centering \includegraphics[width=17cm]{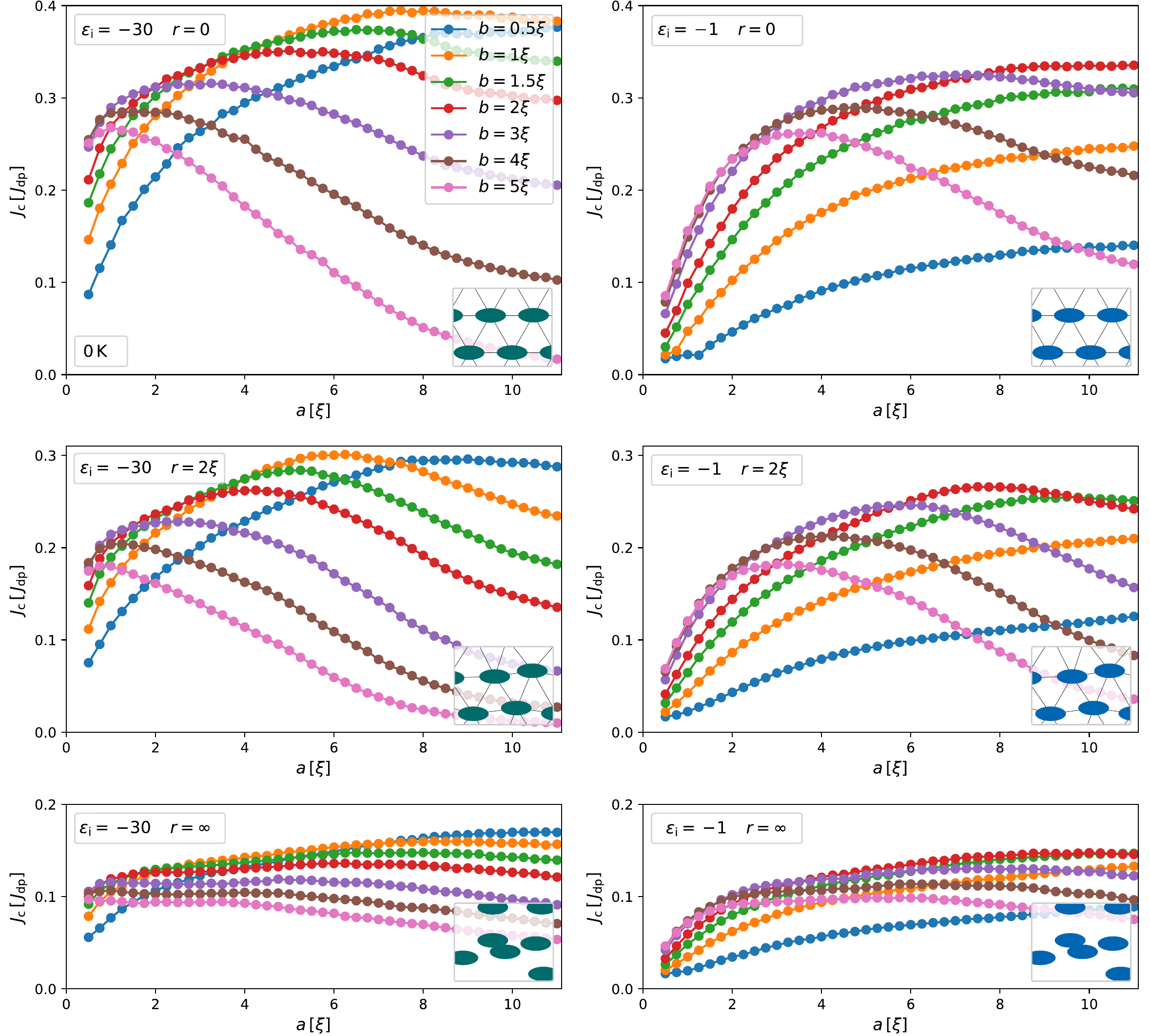}
	\caption{ \label{fig:Jc_a_b_0K}
		Critical current $\Jc$ for hexagonal lattice of columnar defects with elliptical 
		cross section (diameter $a$ is in $x$ direction, $b$ is in $y$ direction) 
		as a function of $a$ for lattice constant $l = 8.5\xi$ and different $b$. 
		Some randomness, $r$, is added to the $x$ and $y$ positions of the defects, 
		i.e. $\delta x, \delta y = [-r, \ldots, r]$. Shown for a perfect hexagonal lattice 
		with $r = 0$ (top row), lattice with intermediate randomness having 
		$r = 2\xi$ (medium row), and uncorrelated placement of defects corresponding 
		to $r = \infty$ (bottom row). 
		Strong $\ei = -30$ (left column) and weak $\ei = -1$ (right column) pinners. 
		A small Langevin noise coefficient is used ($\Tf = 10^{-5}$) corresponding to 
		near-zero temperatures.
	}
\end{figure}

\begin{figure}[ht]
	\centering \includegraphics[width=17cm]{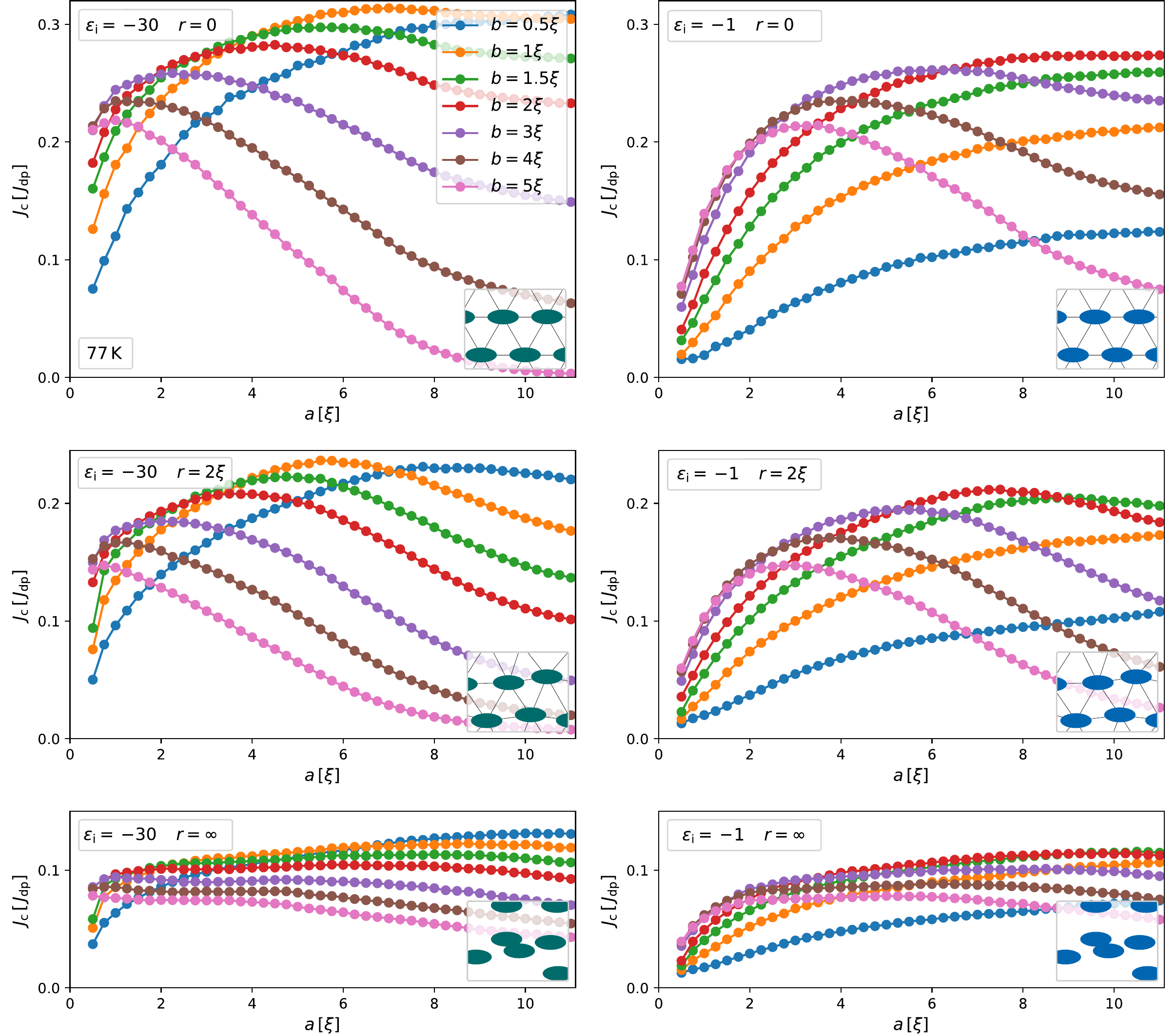}
	\caption{ \label{fig:Jc_a_b_77K}
		The same as in Fig.~\ref{fig:Jc_a_b_0K} but for larger Langevin 
		noise coefficient ($\Tf$ = 0.28) corresponding to 77\,K.
	}
\end{figure}

In Figs.~\ref{fig:Jc_l_b_planar}--\ref{fig:Jc_l_d} we provide a sampling set of $\Jc$ for the hexagonal lattice made of columnar defects with cross section $a$ and $b$ for $a/b$ ratios ranging from planar defects ($a \to \infty$) to cylindrical columns ($a = b$). Figures \ref{fig:Jc_a_b_0K} and \ref{fig:Jc_a_b_77K} demonstrate the influence of the randomness in the defect placement. All plots are for a current applied in $x$ direction and magnetic field of $B = 0.1\Hct$ directed along the $z$ axis.

\end{document}